\documentclass[preprint]{aastex}

\tighten

\input psfig

\def\kmsm{\,{\rm km\, s^{-1}\, Mpc^{-1}} }

\begin{document}
\title{CLASS B0739+366: A new two-image gravitational lens system}

\author{D.R. Marlow, D. Rusin}
\affil{Department of Physics and Astronomy, University of Pennsylvania, 
209 S. 33rd St., Philadelphia, PA 19104-6396}

\author{M. Norbury, N. Jackson, I.W.A. Browne, P.N. Wilkinson}
\affil{NRAL, Jodrell Bank, University of Manchester, Macclesfield, Cheshire
SK11 9DL, UK} 

\author{C.D. Fassnacht, S.T. Myers\altaffilmark{1}}
\affil{National Radio Astronomy Observatory,
P.O. Box 0, Socorro, NM 87801}
\altaffiltext{1}{and Department of Physics and Astronomy, University of
Pennsylvania, 209 S. 33rd St., Philadelphia, PA 19104-6396}

\author{L.V.E. Koopmans\altaffilmark{2}}
\affil{Kapteyn Astronomical Institute, Postbus 800, NL-9700 AV Groningen, 
The Netherlands}
\altaffiltext{2}{and NRAL, Jodrell Bank, University of Manchester,
Macclesfield, Cheshire SK11 9DL, UK} 

\author{R.D. Blandford, T.J. Pearson, A.C.S. Readhead}
\affil{California Institute of Technology, 105-24, Pasadena, CA 91125}

\author{A.G. de Bruyn\altaffilmark{3}}
\affil{NFRA, Postbus 2, NL-7990 AA Dwingeloo, The Netherlands}
\altaffiltext{3}{and Kapteyn Astronomical Institute, 
Postbus 800, NL-9700 AV Groningen, The Netherlands}

\begin{abstract}

We present the discovery of CLASS B0739+366, a new gravitational lens system
from the Cosmic Lens All-Sky Survey. Radio imaging of the source with the Very
Large Array (VLA) shows two compact components separated by $0\farcs54$, with
a flux density ratio of $\sim$ 6:1. High-resolution follow-up observations
using the Very Long Baseline Array (VLBA) at 1.7 GHz detect weak,
parity-reversed jet emission from each of the radio components.  Hubble Space
Telescope NICMOS F160W observations detect infrared counterparts to the lensed
images, as well as an extended object between them which we identify as the
lensing galaxy.  Redshifts for the galaxy and lensed source have not yet been
obtained. For typical lens and source redshifts of $z=0.5$ and $z=1.5$,
respectively, preliminary mass modeling predicts a time delay of $\sim7h^{-1}$
days in a flat $\Omega_{M}=1.0$ universe. The small predicted time delay and
weak radio components will make CLASS B0739+366 a challenging target for
Hubble constant determination.

\end{abstract}


\section{Introduction} \label{sec:intro}

Gravitational lensing is a potent tool for addressing a number of current
problems in cosmology and astrophysics. First, measured time delays between
the images of a lensed source, when combined with a well-constrained mass
model for the system, allow for a determination of the Hubble constant, $H_o$
(Refsdal 1964). Thus far time delays have been measured for six gravitational
lens systems (Schechter et al.\ 1997; Kundic et al.\ 1997; Lovell et al.\
1998; Biggs et al.\ 1999; Fassnacht et al.\ 1999b; Koopmans et al.\ 2000) and
favor a Hubble constant of $H_o = 66 \pm 6 \kmsm$ (Koopmans \& Fassnacht
1999). Second, the lensing rate in a systematic survey can place strong limits
on the cosmological density parameters, particularly the cosmological constant
(Turner, Ostriker \& Gott 1984; Turner 1990). For flat ($\Omega_m +
\Omega_{\Lambda} = 1$) cosmological models, recent analyses favor
$\Omega_{\Lambda} < 0.65$ at $95\%$ confidence (Kochanek 1996; Falco, Kochanek
\& Mu\~noz 1998; Helbig et al.\ 1999). Finally, gravitational lensing is an
excellent means of constraining the properties of galaxy mass distributions
(Kochanek 1991).

The Cosmic Lens All-Sky Survey (CLASS; Browne et al.\ 1999; Myers et al.\ in
preparation) seeks to discover new cases of gravitational lensing among
flat-spectrum radio sources. Such sources are compact, which eases the
identification of multiple imaging, and variable, so time delays can be
measured from component lightcurves.  CLASS builds upon the success of the
Jodrell-VLA Astrometric Survey (JVAS; Patnaik et al.\ 1992; Browne et al.\
1998; Wilkinson et al.\ 1998; King et al.\ 1999) and extends the search to
weaker flux densities. CLASS has observed over $15000$ sources with the Very
Large Array (VLA) in A-configuration at 8.4 GHz. The observations have been
performed over four phases (CLASS 1--4) from spring 1994 to summer 1999, and
are now complete.

Sources in the first two phases of CLASS were selected from the 87GB 5 GHz
catalogue (Gregory \& Condon 1991), with $S_{5}\geq 25$ mJy and spectral
index $\alpha\geq-0.5$ (where $S_{\nu} \propto \nu^{\alpha}$) between 5 GHz
and the 327 MHz Westerbork Northern Sky Survey (WENSS; Rengelink et al.\ 1997)
or the 365 MHz Texas Survey (Douglas et al.\ 1996). Recently, the sources in
the CLASS sample were reselected using the 5 GHz GB6 catalogue (Gregory et
al.\ 1996) and the 1.4 GHz NRAO VLA Sky Survey (NVSS; Condon et al.\ 1998),
with a 5 GHz cutoff of 30 mJy and the same spectral criteria. The final CLASS
statistical sample will consist of $\sim 12000$ sources (Myers et al.\ in
preparation).

The VLA survey data are calibrated in the data reduction package AIPS with the
usual procedure, and are mapped using an automated script within the imaging
package DIFMAP (Shepherd 1997). The data are then modeled with Gaussian
components, which provides a quantitative means of describing the observed
radio morphologies. Sources modeled with multiple compact components are
selected as preliminary lens candidates. These candidates are followed up
with high-resolution radio observations using the Multi-Element Radio-Linked
Interferometer Network (MERLIN) and then the Very Long Baseline Array (VLBA)
for the few surviving sources. The vast majority of the lens candidates are
rejected on surface brightness and morphological grounds. Those candidates
surviving the radio filter are followed up further with optical and/or 
near-infrared imaging and spectroscopy.

The first phase of CLASS observations (CLASS--1) has yielded four new lens
systems: B0712+472 (Jackson et al.\ 1998), B1600+434 (Jackson et al.\ 1995),
B1608+656 (Myers et al.\ 1995) and B1933+503 (Sykes et al.\ 1998).  The second
series of observations (CLASS--2) has until now produced three new lenses:
B1127+385 (Koopmans et al.\ 1999), B1555+375 (Marlow et al.\ 1999) and
B2045+265 (Fassnacht et al.\ 1999a).  Two additional lens systems have
recently been discovered during the third phase of CLASS observations
(CLASS--3): B1152+199 and B1359+154 (Myers et al.\ 1999; Rusin et al.\
2000). Radio follow-up observations of the remaining CLASS--3 and CLASS--4
candidates are currently taking place.

Here we present the discovery of a new two-image gravitational lens system
from CLASS--2: B0739+366. In section 2 we outline the radio observations of
the source. Section 3 describes NICMOS imaging of the lens. In section 4 a
preliminary mass model for the lens system is presented. Section 5 concludes
with a summary of our results and discusses future work.

\section{Radio Observations} \label{sec:obs}

\subsection{VLA and MERLIN Observations}

The B0739+366 system was observed on 1995 August 13 with the VLA at 8.4 GHz in
A-configuration during the second phase of the CLASS survey observations. The
data were modeled within DIFMAP by two compact Gaussian components separated
by $0\farcs54$, with a flux density ratio of 6.4:1.  The final map is shown in
Fig.\ 1 and has an angular resolution of $\sim0\farcs2$.  Follow-up VLA 5, 8.4
and 15 GHz A-configuration observations were performed on 1999 August 30.  The
component flux densities are given in Table 1. The spectral indices of the
components are $\alpha_A = -0.3 \pm 0.1$ and $\alpha_B = -0.5 \pm 0.1$
($S_{\nu}\propto\nu^{\alpha}$), respectively, between 5 GHz and 15 GHz. Since
the B0739+366 VLA survey data were fitted with multiple compact components,
the source was considered a promising lens candidate and selected for
high-resolution radio follow-up.

MERLIN 5 GHz snapshot observations of B0739+366 were performed on 1996
December 27. The data were modeled by two compact components with a flux
density ratio of 6.9:1. Subsequently, a second MERLIN 5 GHz observation was
made on 1998 December 4.  The total integration time was 12 hours. The final
map (Fig.\ 2) shows a slight extension to the primary image. The compactness
of the components and their similarity in spectral index are evidence of the
lensed nature of this system.

\subsection{VLBA Observations}

VLBA 5 GHz observations of B0739+366 were obtained on 1997 August 2. Snapshots
were taken over a range of hour angles to synthesize $uv$-coverage. The total
integration time was 32 minutes.  Phase referencing was performed by switching
from the nearby strong VLBI calibrator B0738+313. Fringe-fitting was performed
on B0738+313 and the solutions were transferred to B0739+366 directly. Both
components seen in the MERLIN map were detected. The naturally-weighted 5 GHz
VLBA maps are shown in Fig.\ 3 and have a resolution of $3.8 \times 1.5$ mas
and an RMS noise level of $110$ $\mu$Jy/beam. The primary component shows
evidence of substructure and was modeled by two Gaussian subcomponents. The
secondary component appears to be very slightly resolved, but was modeled by a
single Gaussian. The positions, flux densities and sizes of the VLBA modelfit
components are listed in Table 2.

A deep 1.7 GHz VLBA observation of the system was obtained on 1999 August 29.
Alternate observations of the target source and a phase reference source
(B0749+376) were iterated on a cycle of 3 minutes and 1.5 minutes,
respectively. The total integration time was 8 hours. Fringe-fitting was
performed on B0749+376 and the solutions were transferred to
B0739+366. Naturally-weighted 1.7 GHz VLBA longtrack maps of B0739+366 are
shown in Fig.\ 4. The maps have a resolution of $7.7 \times 5.4$ mas and an
RMS noise level of $45$ $\mu$Jy/beam. Weak, extended jet emission is detected
to the east of component A and to the west of component B. The
oppositely directed jets in A and B are characteristic of parity reversal
induced by gravitational lensing.

\section{HST/NICMOS Observations}

Hubble Space Telescope (HST) observations of B0739+366 were performed on
1998 April 10 with the Near-Infrared Camera and Multi-Object Spectrometer
(NICMOS).  The F160W filter was used, which corresponds roughly to 
ground-based $H$ band. The observations were performed with the NIC1 camera,
which has a detector scale of 43 mas/pixel and a field-of-view of
$11\farcs0\times 11\farcs0$. The total exposure time was 2624 seconds. 
The data were subjected to the standard calibration pipeline involving bias
and dark current subtraction, linearity and flat-field correction, photometric
calibration and cosmic ray removal. 

The calibrated image (Fig.\ 5) shows two clear emission peaks corresponding
to the lensed images. The ring surrounding component A is due to the NICMOS
point-spread function (PSF). In addition, an extended emission feature appears
to be blended with component B. We identify this with the lensing galaxy.  The
relative positions of the lensed images match those seen at radio wavelengths
to within $10$ mas.  The F160W magnitudes are $m_A = 19.0$ for image A, $m_B =
21.7$ for image B and $m_{gal} = 21.7$ for the galaxy. 

A PSF subtraction was performed to better study the lensing galaxy. First a
standard TinyTim PSF (Krist \& Hook 1997) for the NICMOS/F160W filter was
fitted to the center of component A using the MAXFIT task in the AIPS data
analysis package. Component A was removed using a manually-scaled best-fit
PSF. The PSF was then shifted to the radio position of component B, which was
removed using a scaled fit. The ratio subtracted at A and B was 6.3 with an
error of approximately 0.4. This is very close to the ratio of flux densities
measured for the radio components, and indicates very little dust extinction
at $H$ band, assuming no variability of the lensed source. The subtracted
image (Fig.\ 5) offers a greatly improved view of the lensing galaxy. The
detection of optical counterparts to the radio components along with extended
galaxy emission provides strong evidence for the lensing hypothesis.

\section{Modeling} \label{sec:sys}

In this section we present a preliminary mass model that describes the
observed properties of the B0739+366 lens system. We assumed a flat
$\Omega_{M}=1$ universe with $H_{0} = 100h \kmsm$, a lens redshift of $z=0.5$
and a source redshift of $z=1.5$ for all calculations.

We modeled the system with a singular isothermal ellipsoid mass distribution
(SIE; Kormann, Schneider \& Bartelmann 1994). Although each radio component
exhibits core and jet emission, the jet features are quite weak and highly
resolved. Since their centroids are poorly determined, they are unlikely to 
provide good additional modeling constraints at the present time. Therefore we
use only the positions of the core components in our modeling analysis. 

To limit the number of free parameters and ensure a constrained model, we
fixed the SIE on the surface brightness center of the lens galaxy from the
NICMOS image. The model therefore has only five free parameters -- the lens
velocity dispersion, axial ratio and position angle, and two source
coordinates. The 5 GHz VLBA data provided five constraints -- four absolute
image coordinates and a flux density ratio. The number of degrees of freedom
(NDF) is therefore zero. The position of component A was taken to be that of
VLBA component A1. The flux density of component A was taken to be the sum of
A1 and A2.

Modeling was performed with an image plane minimization, as described by
Kochanek (1991). We optimized the goodness of fit parameter using the 
image positions $\theta_{i}$ and magnification ratio $r=|S(B)/S(A)|$, 
\begin{equation}
\chi^{2} = \sum_{i=A,B} \left[ \frac{(\theta_{i}' - 
\theta_{i})^{2}}{\Delta \theta_{i}^{2}}\right] +
\frac{(r' -r )^{2}}{\Delta r^{2}}
\end{equation}
where primed quantities are model-predicted and unprimed quantities are
observed. Because NDF $=0$, the best-fit model is described by $\chi^2 = 0$,
independent of the positional and flux density errors. The optimized model
parameters, predicted magnifications and time delays are listed in Table
3. The critical curves and caustics for the system are shown in Fig.\ 6. 

A series of Monte Carlo simulations was performed to examine the stability of
the best-fit model parameters. Gaussian-distributed errors accounting for
observational uncertainties were added simultaneously to the galaxy position
($10$ mas in each coordinate) and component flux densities ($10\%$). This
latter uncertainty also accounts for the possible effect of source variability
on the flux density ratio.  The optimized model parameters and recovered time
delay were determined for each of 10,000 trials. The uncertainties in the
best-fit model parameters were found from the ranges enclosing $95\%$ of the
results, and are given in Table 3.

\section{Summary and Future Work} \label{sec:fut}

We have discovered a new gravitational lens system in the Cosmic Lens All-Sky
Survey -- CLASS B0739+366. The system consists of two components separated by
$0\farcs54$, with a flux density ratio of $\sim$ 6:1. VLA observations have
determined that the spectral indices of the radio components agree within the
measurement errors. Follow-up observations with the VLBA have detected jet
emission extending from each of the radio components. The jets are oppositely
directed, and provide a classic example of lensing-induced parity reversal.
All available radio data of B0739+366 is therefore consistent with what is
expected for a gravitational lens system.

HST/NICMOS imaging has detected infrared counterparts to each of the 
radio components, as well as an extended emission feature which we identify
with the lensing galaxy. These observations offer further proof for the
lensing hypothesis. We note that no lens candidate that has survived
high-resolution radio and optical filters such as those presented here has
ever been rejected by additional observations. 

A preliminary lens model has been constructed using constraints derived from
the VLBA and HST data. Assuming typical lens and source redshifts of $z=0.5$
and $z=1.5$, respectively, the predicted time delay between the lensed
components is $\sim7h^{-1}$ days for $H_{0}=100h \kmsm$ and a flat
$\Omega_{M}=1.0$ cosmology. The error on the time delay due to observational
uncertainties is roughly $10\%$.

Deeper VLBA or global VLBI observations of B0739+366 are required to study the
extended jet emission present in each of the lensed radio components, with a
view to acquiring additional constraints on our lens model. Such observations
would demand significantly longer integrations than those already undertaken
($\sim 8$ hr), as the jets are very weak and extended. Future optical
observations will focus on obtaining redshifts for both the lensing galaxy and
lensed source. Redshifts are necessary to convert a measured time delay into a
value of the Hubble constant, and will also provide important input to
statistical analyses of the CLASS survey data.

Even if the redshifts can be measured, CLASS B0739+366 will be a challenging
lens system to use for Hubble constant determination. First, the predicted
time delay is quite short, which would require a rather densely sampled
monitoring program. Second, the weakness of component B will make it difficult
to construct a clean lightcurve for this image.  A moderately variable source
(with an amplitude of $\sim 10\%$) would produce features at the level of only
a few tenths of a mJy in component B, which is approximately the error
associated with absolute flux densities derived from VLA monitoring
observations (eg.\ Fassnacht et al.\ 1999b; Koopmans et al.\ 2000). In
addition, the lightcurves of weak compact radio images could be susceptible to
scintillation, as the component sizes may not be more than a few tens of
microarcseconds (eg.\ Koopmans \& de Bruyn 2000). For these reasons B0739+366
is not considered one of the best CLASS lenses with which to determine
$H_o$. However, when combined with the other lenses discovered in the JVAS and
CLASS surveys, this system will add to the statistical evidence placing limits
on the values of the cosmological energy densities.

\acknowledgments

We thank the staffs of the VLA, MERLIN and VLBA for their assistance during
our observing runs. We also thank the anonymous referee, whose comments and
suggestions led to a greatly improved draft. The National Radio Astronomy
Observatory is a facility of the National Science Foundation operated under
cooperative agreement by Associated Universities, Inc. MERLIN is operated as a
National Facility by NRAL, University of Manchester, on behalf of the UK
Particle Physics \& Astronomy Research Council. This research used
observations with the Hubble Space Telescope, obtained at the Space Telescope
Science Institute, which is operated by Associated Universities for Research
in Astronomy Inc. under NASA contract NAS5-26555.  DR gratefully acknowledges
a fellowship from the Zaccheus Daniel Foundation. STM was supported by an
Alfred P. Sloan Fellowship at the University of Pennsylvania. This research
was supported in part by European Commission TMR Programme, Research Network
Contract ERBFMRXCT96-0034 ``CERES''.

\clearpage

\begin{table} \begin{tabular}{ c c c c c c c c c c} \hline \hline Comp &
$S_{5}$ (mJy) & $S_{8.4}^{1}$ (mJy)& $S_{8.4}$ (mJy)  & $S_{15}$ (mJy) \\
\hline A &26.8 & 21.7&20.3 &19.1 \\ B &4.9 & 3.4 &3.8 &2.8 \\ \hline
\end{tabular} \caption{CLASS B0739+366 VLA component flux densities. Data
for the 1999 August 30 observations, except (1) the VLA 8.4 GHz survey
observation of 1995 August 13. Errors on the flux densities are $\sim$10
\%.} \end{table}

\begin{table}
\begin{tabular}{ c c c c c  }
\hline 
Comp & \sc Offset (E) & \sc Offset (N)   &$S_{5}$ (mJy) & $b_{maj}$ (mas) \\ 
\hline
A1 & 0 & 0 &  29.0 & 1.6\\
A2 & $+0\farcs0024$  & $+0\farcs0010$    &  2.1  & 1.7   \\
B  & $+0\farcs2217$  & $-0\farcs4910$     &  6.2 & 2.0   \\
\hline 
\end{tabular}
\caption{CLASS B0739+366 VLBA 5 GHz modelfit component positions, flux
densities and major axes. Data for the 1997 August 2 snapshot
observation. Positions are offset from RA 07 42 51.1685 Dec +36 34 43.638
(J2000.0). Errors on the positions are 0.1 mas. Errors on the flux densities
are $\sim$10 \%.}
\end{table}

\begin{table}
\begin{tabular}{cc}
\hline\hline
Parameter & Model \\
\hline
$(b/a)_G$  & $0.77_{-0.30}^{+0.17}$\\
$\sigma_{G}$ (km/s)  &$135.1_{-2.5}^{+3.4}$\\
$PA_{G}^{\circ}$  & $87.3_{-39.2}^{+31.7}$\\
$(x,y)_{G}$     & ($+0\farcs1840$, $-0\farcs4324$)\\
$(x,y)_{src}$  &
($+0\farcs0929^{+0\farcs0352}_{-0\farcs0291}$,
$-0\farcs2550^{+0\farcs0161}_{-0\farcs0236}$)\\
$\mu_{A,B}$ & $+2.01$, $-0.40$\\
$\Delta t$ ($h^{-1}\ $ days) & $+6.64_{-0.64}^{+0.65}$\\
\hline
\end{tabular}

\caption{CLASS B0739+366 best-fit SIE mass model parameters. Listed are the
axial ratio $(b/a)_G$, velocity dispersion ($\sigma_{G}$) and position angle
($PA_{G}$) of the SIE. The fixed position of the lensing galaxy is given by
$(x,y)_{G}$, as derived from the NICMOS image. The observational error on this
position is $\sim 0\farcs01$.  The recovered source position is denoted by
$(x,y)_{src}$. The magnifications of the lensed images are given by
$\mu_{A,B}$. The predicted time delay is $\Delta t$. The errors on all model
parameters indicate the $95\%$ confidence interval determined from Monte Carlo
simulations.}
\end{table}

\clearpage

\begin{figure*}
\figurenum{1}
\psfig{file=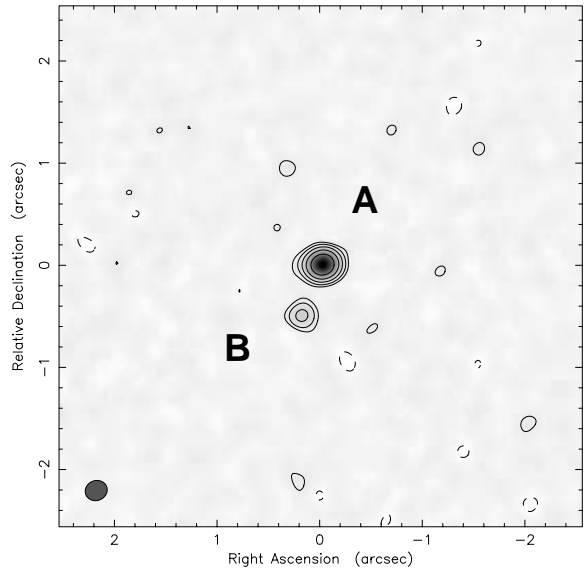,width=3.0in}
\caption{VLA 8.4 GHz snapshot map of CLASS B0739+366 taken 1995
August 13. The image has a resolution of $0\farcs24 \times 0\farcs21$ (PA
$-65.6^{\circ}$).} 
\end{figure*}

\begin{figure*}
\figurenum{2}
\psfig{file=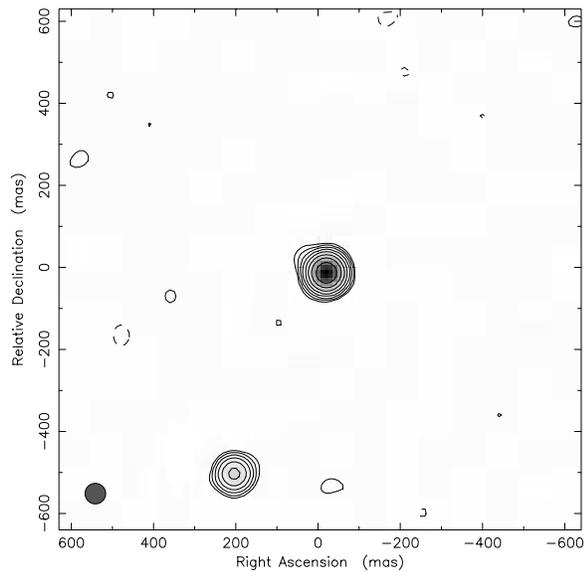,width=3.0in}
\caption{MERLIN 5 GHz longtrack map of CLASS B0739+366
taken 1998 December 4. The image is restored with a 50 mas circular beam.}
\end{figure*}

\clearpage

\begin{figure*}
\begin{tabular}{c c}
\psfig{file=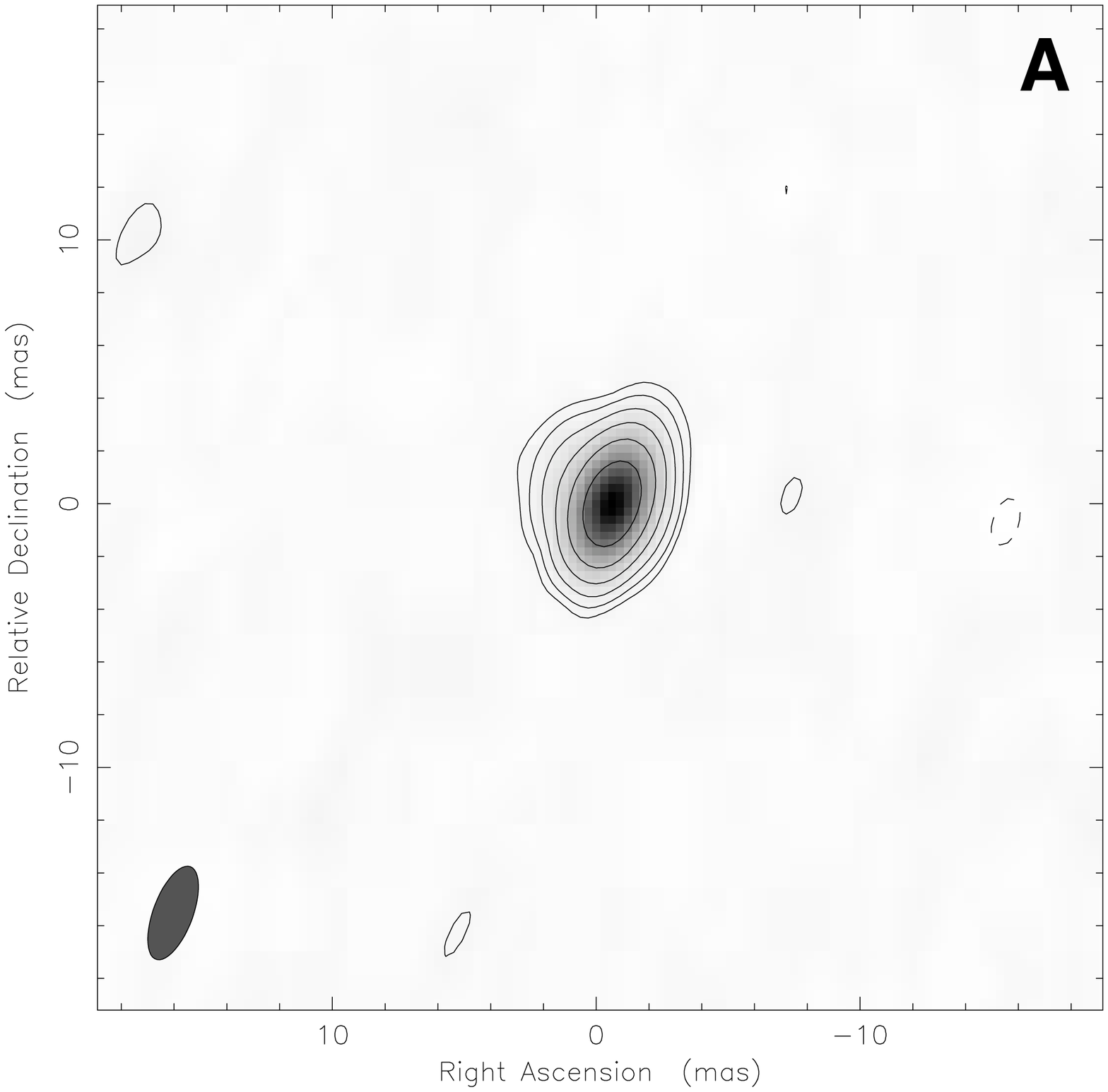,width=3.0in} & 
\psfig{file=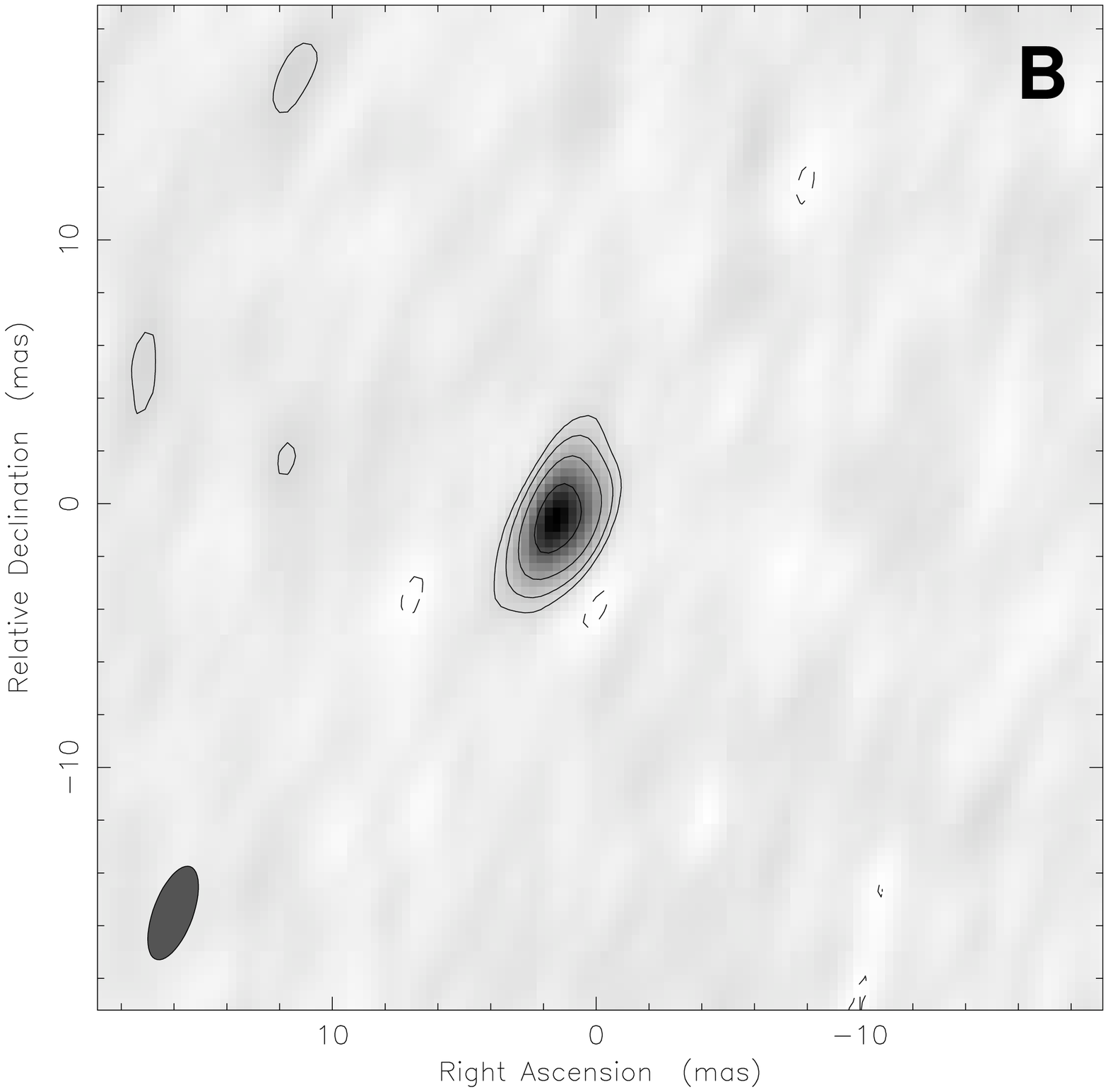,width=3.0in}\\
\end{tabular}
\figurenum{3}
\caption{VLBA 5 GHz snapshot maps of CLASS B0739+366 taken 1997 August 2.
Maps have a resolution of $3.8 \times 1.5$ mas (PA $-20.4^{\circ}$).  (a) Left:
Component A. (b) Right: Component B.}
\end{figure*}

\clearpage

\begin{figure*}
\begin{tabular}{c c}
\psfig{file=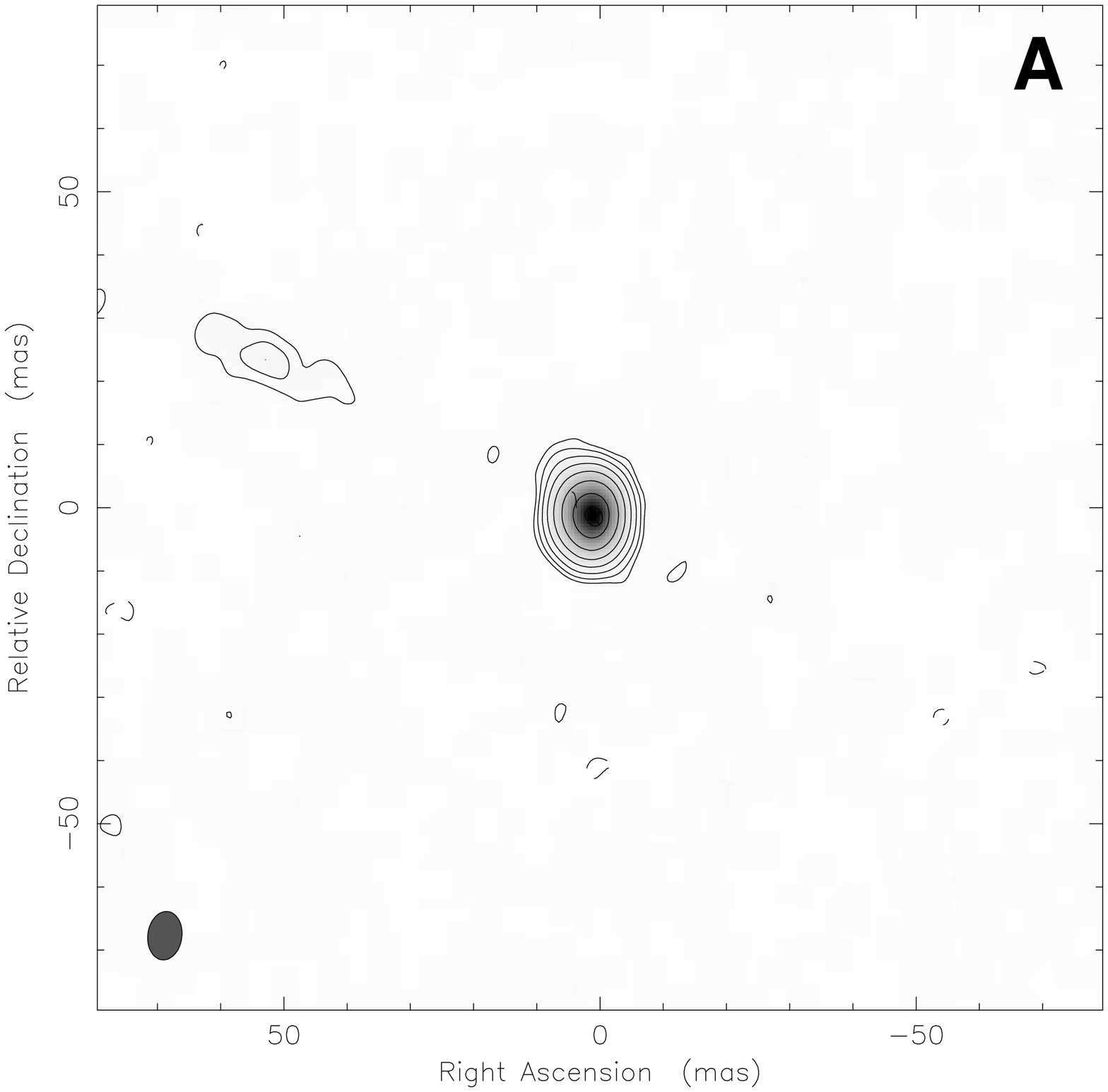,width=3.0in} & 
\psfig{file=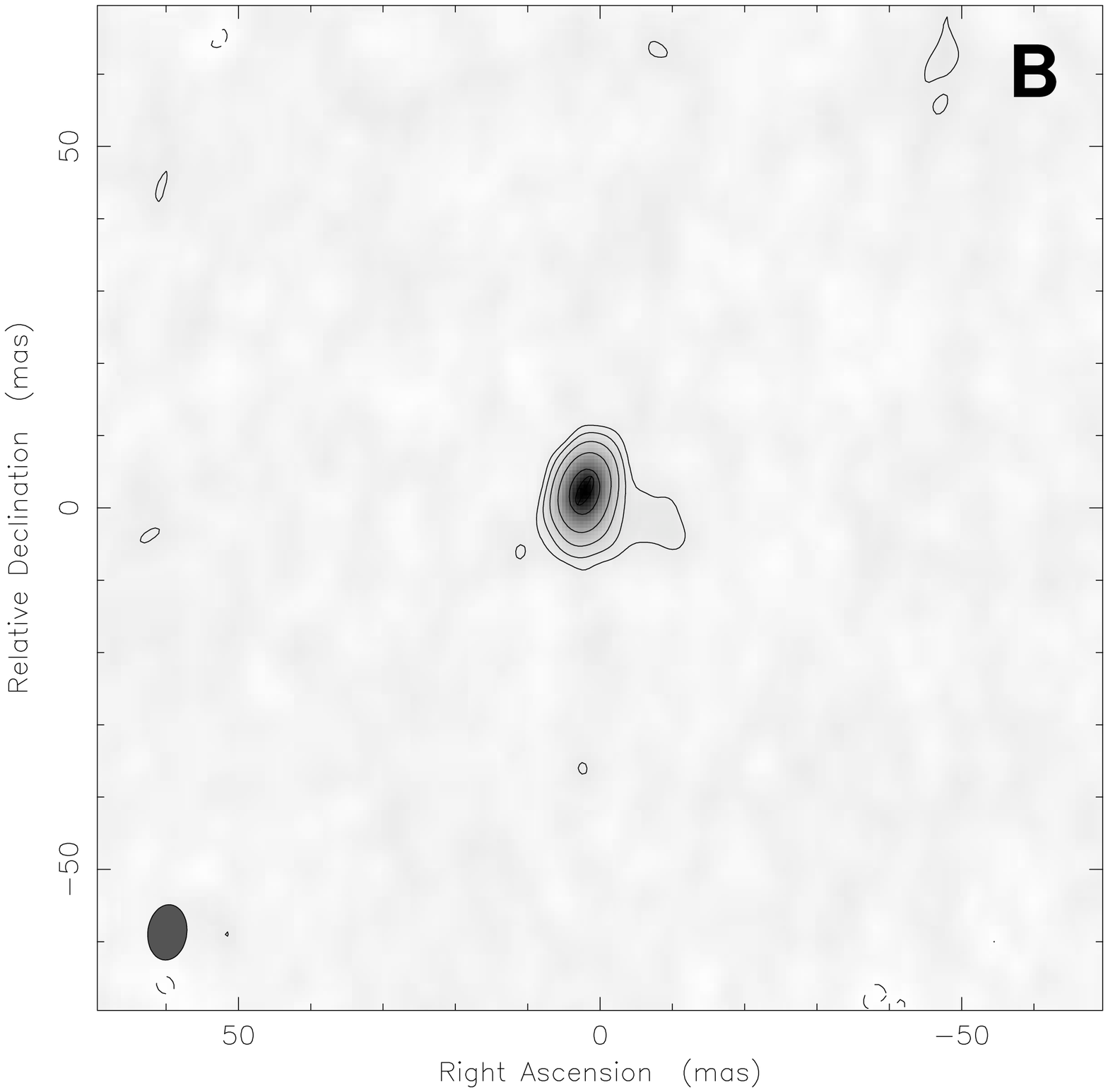,width=3.0in}\\
\end{tabular}
\figurenum{4}
\caption{VLBA 1.7 GHz longtrack maps of CLASS B0739+366 taken 1999 August 29.
Maps have a resolution of $7.7 \times 5.4$ mas (PA $-7.1^{\circ}$). (a) Left:
Component A. (b) Right: Component B.}
\end{figure*}

\clearpage 

\begin{figure}
\psfig{file=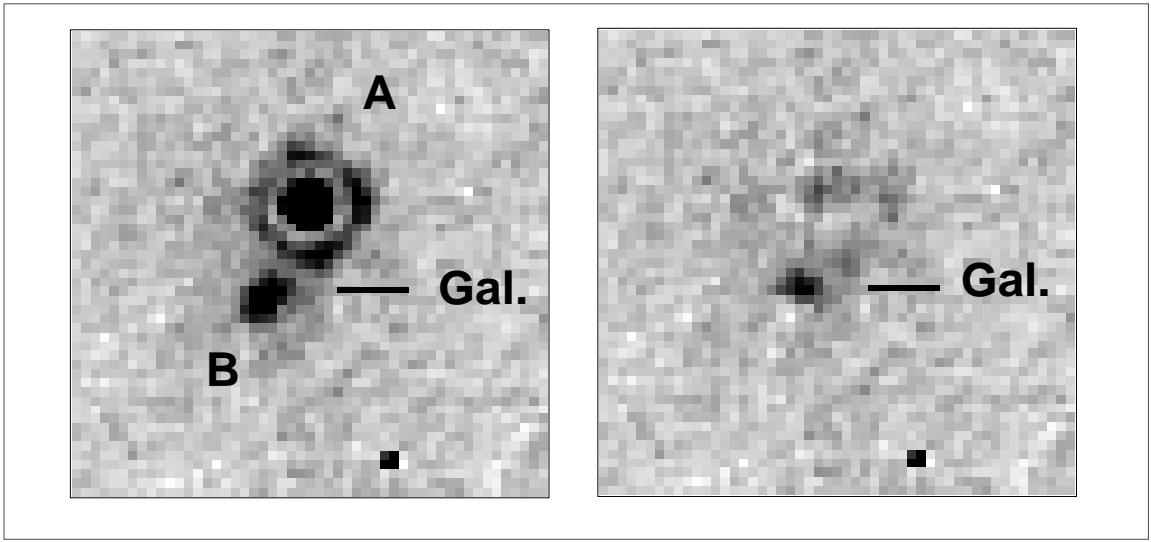,width=6in}
\figurenum{5}
\caption{NICMOS F160W image of CLASS B0739+366. Left: Two peaks of optical
emission correspond to the lensed images A and B. The lensing galaxy appears
as an extension to component B. North is up, east is left. The image size is
$3\farcs7 \times 3\farcs7$. Right: PSF-subtracted image brings out emission
from the lensing galaxy.}
\end{figure}

\clearpage

\begin{figure}
\psfig{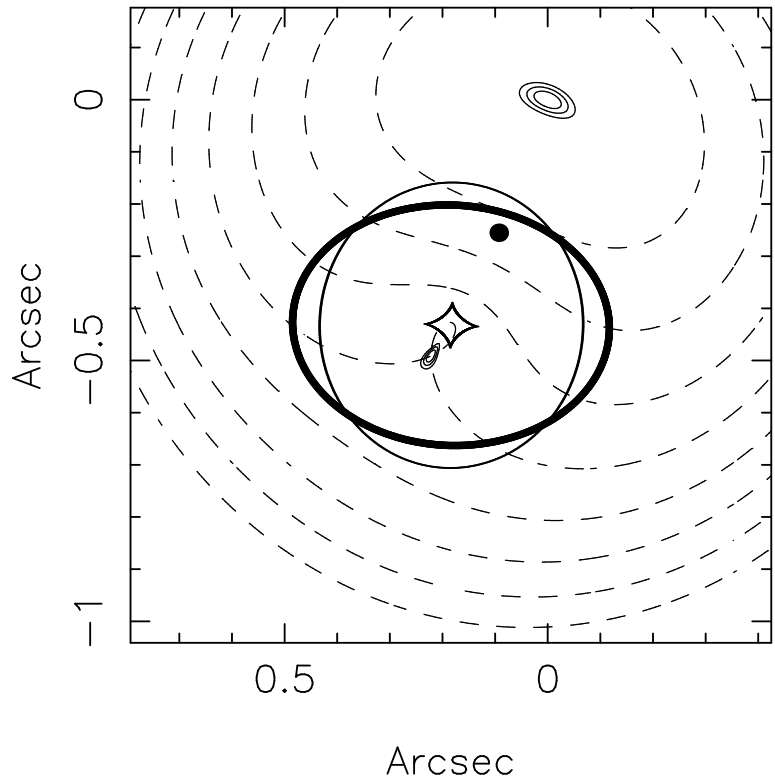}
\figurenum{6}
\caption{Lens model for CLASS B0739+366. The critical curves (dark line) and
caustics (light lines) are shown for the SIE model. The position of the source
is marked by a filled circle. The solid contours describe the shape of the
radio components if the source surface brightness distribution is
Gaussian. The dashed contours mark curves of constant time delay, in
increments of $1.7$ $h^{-1}$ days outward from component A. }

\end{figure}

\end{document}